\documentclass[twocolumn,aps,amsmath,amssymb,floatfix,superscriptaddress,prb,longbibliography]{revtex4-1}
\usepackage{graphicx}
\usepackage{dcolumn}
\usepackage{bm}
\usepackage{natbib}
\usepackage{xspace}
\usepackage{hyperref}
\usepackage{color}

\usepackage[flushleft]{threeparttable}
\def\sms{Sr$_{1-x}$Mn$_{1-y}$Sb$_2$\xspace}
\begin{document}

\title{Spin dynamics of a magnetic Weyl semimetal \sms}

\author{Zhengwei Cai}
\altaffiliation{These authors contributed equally to the work.}
\author{Song Bao}
\altaffiliation{These authors contributed equally to the work.}
\author{Wei Wang}
\altaffiliation{These authors contributed equally to the work.}
\author{Zhen Ma}
\affiliation{National Laboratory of Solid State Microstructures and Department of Physics, Nanjing University, Nanjing 210093, China}
\author{Zhao-Yang~Dong}
\affiliation{Department of Applied Physics, Nanjing University of Science and Technology, Nanjing 210094, China}
\author{Yanyan~Shangguan}
\author{Jinghui~Wang}
\author{Kejing Ran}
\author{Shichao Li}
\affiliation{National Laboratory of Solid State Microstructures and Department of Physics, Nanjing University, Nanjing 210093, China}
\author{Kazuya Kamazawa}
\affiliation{Neutron Science and Technology Center, Comprehensive Research Organization for Science and Society (CROSS), Tokai, Ibaraki 319-1106, Japan}
\author{Mitsutaka Nakamura}
\affiliation{J-PARC Center, Japan Atomic Energy Agency (JAEA), Tokai, Ibaraki 319-1195, Japan}
\author{Devashibhai Adroja}
\affiliation{ISIS Facility, Rutherford Appleton Laboratory, Chilton, Didcot, Oxon OX11 0QX, United Kingdom}
\affiliation{Highly Correlated Matter Research Group, Physics Department, University of Johannesburg, P.O. Box 524,
Auckland Park 2006, South Africa}
\author{Shun-Li~Yu}
\email{slyu@nju.edu.cn}
\author{Jian-Xin Li}
\email{jxli@nju.edu.cn}
\author{Jinsheng Wen}
\altaffiliation{jwen@nju.edu.cn}
\affiliation{National Laboratory of Solid State Microstructures and Department of Physics, Nanjing University, Nanjing 210093, China}
\affiliation{Collaborative Innovation Center of Advanced Microstructures, Nanjing University, Nanjing 210093, China}

\begin{abstract}
Dirac matters provide a platform for exploring the interplay of their carriers with other quantum phenomena. \sms has been proposed to be a magnetic Weyl semimetal and provides an excellent platform to study the coupling between Weyl fermions and magnons. Here, we report comprehensive inelastic neutron scattering (INS) measurements on single crystals of \sms, which have been well characterized by magnetization and magnetotransport measurements, both of which demonstrate that the material is a topologically nontrivial semimetal. The INS spectra clearly show a spin gap of $\sim6$~meV. The dispersion in the magnetic Mn layer extends up to about 76~meV, while that between the layers has a narrow band width of 6~meV. We find that the linear spin-wave theory using a Heisenberg spin Hamiltonian can reproduce the experimental spectra with the following parameters: a nearest-neighbor ($SJ_1\sim28.0$~meV) and next-nearest-neighbor in-plane exchange interaction ($SJ_2\sim9.3$~meV) , interlayer exchange coupling ($SJ_c\sim-0.1$~meV), and spin anisotropy constant ($SD\sim-0.07$~meV). Despite the coexistence of Weyl fermions and magnons, we find no clear evidence that the magnetic dynamics are influenced by the Weyl fermions in \sms, possibly because that the Weyl fermions and magnons reside in the Sb and Mn layers separately, and the interlayer coupling is weak due to the quasi-two-dimensional nature of the material, as also evident from the small $SJ_c$ of -0.1~meV.
\end{abstract}

\maketitle

\section{Introduction}

Dirac semimetals are featured by two doubly-degenerate linear bands crossing at the Dirac point\cite{RevModPhys.90.015001,wehling2014dirac,annurev-conmatphys-031113-133841}. The low-energy quasiparticle excitations near the Dirac point can be described by the $4\times4$ Dirac equation. In recent years, they have attracted a lot of attention because of their numerous exotic physical properties, such as high mobility\cite{RevModPhys.81.109,Liang2015,Shekhar2015}, nontrivial $\pi$ Berry phase\cite{Zhang2005,PhysRevLett.113.246402,LI2018535}, and chiral-anomaly-induced negative magnetoresistance\cite{LI2018535,xiong2015evidence,np12_550}. Magnetic Dirac semimetals are particularly intriguing as they not only exhibit these fascinating phenomena but also offer a platform for investigating the interplay between Dirac fermions and spin dynamics. In this regard, the antiferromagnetic 112-type ternary compounds AMnPn$_2$ (A represents alkali- or rare-earth elements such as Sr, Ca, Ba, Yb, or Eu, and Pn = Sb, Bi) have been studied extensively\cite{doi:10.1146/annurev-matsci-070218-010114}. SrMnBi$_2$ was initially reported to be such a material\cite{PhysRevLett.107.126402,PhysRevB.84.064428,PhysRevB.84.220401,PhysRevB.87.245104,sr4_5385,PhysRevB.90.035133,PhysRevLett.113.156602}. It was shown that the linearly dispersing $p$ bands of the Bi atoms formed highly anisotropic two-dimensional Dirac structure near the Fermi level, while the Mn bands were well below the Fermi level with binding energy larger than  2~eV and thus were  localized\cite{PhysRevLett.107.126402,PhysRevB.84.064428,PhysRevB.87.245104,sr4_5385,PhysRevB.90.035133,PhysRevLett.113.156602}. The localized divalent Mn (3$d^5$, spin $S=5/2$) with a magnetic moment of  magnitude $\sim$3.7$\mu_{\rm B}$ formed an antiferromagnetic structure below the N\'{e}el temperature $T_{\rm N}$ of $\sim$300~K\cite{PhysRevB.90.075120,PhysRevB.95.134405}. Similar results with Dirac fermions and magnons in the Bi- and Mn-square layers respectively were found by replacing Sr with Ca (Refs~\onlinecite{PhysRevB.90.075120,PhysRevB.85.041101,PhysRevB.87.245104,PhysRevB.95.134405,doi:10.1063/1.3694760,sr4_5385}), Ba (Refs~\onlinecite{PhysRevB.93.115141,Wang_2016}), Yb~(Refs~\onlinecite{nc10_3424,PhysRevB.94.165161,nc8_646,PhysRevB.94.245101,PhysRevB.96.075151}), or Eu~(Refs~\onlinecite{PhysRevB.90.075109,Masudae1501117,PhysRevB.98.161108}). However, in the AMnBi$_2$ materials, the strong spin-orbital coupling (SOC) of Bi atoms might induce a gap that prevented the formation of the Dirac point, and made the fermions massive\cite{PhysRevLett.107.126402,PhysRevB.96.075151,PhysRevB.100.144431}. One way to avoid the SOC-induced electronic gap opening and preserve the Dirac band structure is to replace Bi with lighter elements that have weaker SOC, {\it e.g.}, Sb~(Ref.~\onlinecite{Farhan_2014}). As a consequence, a number of AMnSb$_2$-type materials where A represents Ca~(Ref.~\onlinecite{PhysRevB.95.045128}), Sr~(Refs~\onlinecite{nm16_905,SciPostPhys.4.2.010,PhysRevB.99.054435,PhysRevB.100.205105}), Ba~(Refs~\onlinecite{Farhan_2014,sr6_30525,Huang6256}), or Yb~(Refs~\onlinecite{PhysRevB.97.045109,PhysRevMaterials.2.021201}) have been discovered.

\begin{figure*}[ht]
\centering
\includegraphics[width=0.95\linewidth]{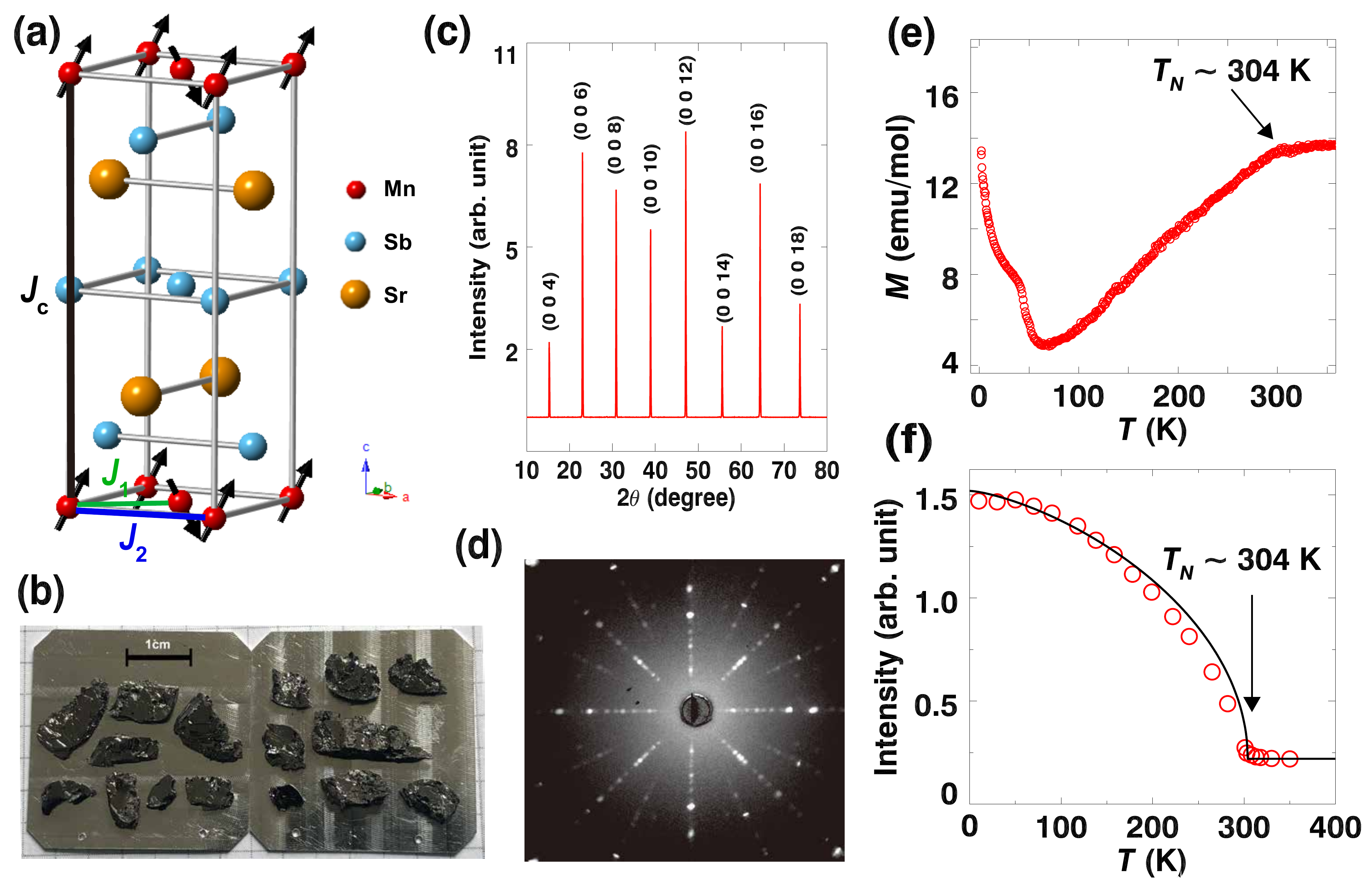}
\caption{(a) Crystal and magnetic structure of \sms (space group No.~62, $Pnma$) with $a \approx b=4.4342$ , $c =$ 23.1359~\AA, and $\alpha = \beta = \gamma = 90^{\circ}$. Note that we only show half of the unit cell along the $c$ axis. Arrows represent the spins. The exchange paths are labeled with $J_1$ (nearest-neighbor), $J_2$ (next-nearest-neighbor),  and $J_c$ (interlayer). (b) Coaligned single crystals we used for the inelastic neutron scattering measurements. (c) Single-crystal X-ray diffraction pattern of the cleavage plane. (d) Backscattering Laue x-ray diffraction pattern with the beam along the [001] direction. (e) Temperature dependence of the magnetization measured with magnetic field $\emph{B}=1$~T applied along the $c$ axis. (f) Temperature dependence of the integrated intensities of the magnetic Bragg peak (1,\,0,\,0). The solid line in (f) is a guideline to the eye. Arrows in (e) and (f) indicate the N\'{e}el temperature $T_{\rm N}$.
\label{fig1}}
\end{figure*}

Among these materials, \sms with Sr and Mn deficiencies  stood out. It was suggested to be a magnetic Weyl semimetal which breaks the time-reversal symmetry\cite{PhysRevB.83.205101,nm16_905,Maeaaw4718}, as opposed to non-magnetic ones that break the spatial inversion symmetry\cite{PhysRevX.5.011029,xu2015discovery,lv2015observation}. Similar to AMnBi$_2$-type magnetic Dirac materials, \sms was shown to exhibit fascinating properties that originated from the two-dimensional Sb and Mn square nets [Fig.~\ref{fig1}(a)]~(Ref.~\onlinecite{nm16_905}). Furthermore, the reduced size of the SOC because of the lighter mass of Sb as compared to Bi resulted in a nearly massless topological fermion behavior ($m^*=0.04-0.05m_0$, where $m^*$ and $m_0$ are the effective and free-electron masses, respectively)\cite{nm16_905}. More importantly, it was shown that below the $T_{\rm N}$ of 304~K, the material underwent a transition to a canted antiferromagnetic phase with a ferromagnetic component of 0.2~$\mu_{\rm B}$~(Ref.~\onlinecite{nm16_905}). This small but finite ferromagnetic moment was proposed to break the time-reversal symmetry, lift the degeneracy of the electron bands, and lead to the long-sought magnetic Weyl semimetal state\cite{PhysRevB.83.205101,nm16_905,nc10_3424}.

However, despite extensive studies on the aforementioned materials, whether and how the topological ferminons interplay with the magnons are still unsettled issues. In CaMnBi$_2$, it was shown that there was a weak resistivity anomaly at the $T{\rm_N}$, indicative of a coupling between the Dirac bands and the magnetic ground state\cite{PhysRevB.90.075120}, but no such anomaly was found in other reports\cite{PhysRevB.85.041101,doi:10.1063/1.3694760}. A two-magnon Raman scattering study of CaMnBi$_2$ and SrMnBi$_2$ indicated that Dirac carriers significantly enhanced the exchange coupling between the magnetic layers ($J_c$), and a spin-fermion model showed that this in turn drove a charge-gap opening along the Dirac locus\cite{nc7_13833}. However, an inelastic neutron scattering (INS) study suggested that the $J_c$ estimated from the Raman measurements might have been exaggerated due to the neglecting of the single-ion spin anisotropy\cite{nc7_13833}, and instead it was found that the Dirac fermions did not affect the spin dynamics at all\cite{PhysRevB.95.134405}. In YbMnBi$_2$, nearly wave-vector independent damping in the INS spectra was taken as a signature of the spin-fermion coupling\cite{arXiv:1908.08114}. In contrast, another INS work did not detect any anomalous features in the magnetic spectra, implying the absence of the coupling between magnons and topological fermions\cite{PhysRevB.100.144431}.

In this work, we have carried out INS together with magnetization and magnetotransport measurements on single crystals of \sms. We observe strong de Haas van Alphen (dHvA) and Shubnikov de Hass (SdH) quantum oscillations at 2~K, from which we find topological fermions with $m^*=0.10m_0$ and a nontrivial Berry phase, consistent with this material being a magnetic Weyl semimetal\cite{nm16_905}. Our INS study reveals clear magnetic spectra dispersing from 6~meV and extending up to $\sim$76~meV. The spin-wave spectra can be well described by a localized Heisenberg model with a nearest-neighbor- ($SJ_1\sim28.0$~meV) and next-nearest-neighbor in-plane exchange interaction ($SJ_2\sim9.3$~meV), and interlayer  exchange coupling ($SJ_c\sim-0.1$~meV), along with a spin anisotropy term ($SD\sim-0.07$~meV). From our results, we do not observe signatures that the magnetic dynamics are affected by the topological fermions.

\section{Experimental Details}

Single crystals of \sms were grown using Sb as the self flux, with an initial atomic ratio of Sr:Mn:Sb = 1:1:4 for the raw materials. An alumina crucible containing the raw materials was sealed under Ar atmosphere in a quartz tube. The sealed ampule was heated up to 850~$^{\circ}$C, followed by a slow cooling at a rate of 1~$^{\circ}$C/h and centrifuged at 660~$^{\circ}$C. Single crystals with size up to $7\times5\times3$~mm$^3$ were so obtained, and some of the crystals are shown in Fig.~\ref{fig1}(b). The composition of the single crystals was measured with energy dispersive spectroscopy (EDS) in an energy analyser (EX-250, Horiba) equipped in a scanning tunneling microscope (S-3400N II, Hitachi). Orientations of the crystals were determined by a single-crystal X-ray diffractometer (D8 Discover, Bruker) and a backscattering Laue X-ray diffractometer (LAUESYS, Photonic Science). Magnetization and resistivity measurements were conducted in a physical property measurement system (PPMS-9T, Quantum Design).

For INS experiments, single crystals were glued on aluminum plates by hydrogen-free Cytop grease. These crystals were well coaligned using a backscattering Laue X-ray diffractometer. Time-of-flight INS experiments were performed on 4SEASONS located at J-PARC Center in Japan\cite{doi:10.1143/JPSJS.80SB.SB025}, and MERLIN at ISIS Facility in the United Kingdom. We coaligned 27 pieces of single crystals weighed about 3.6~g in total for the experiment on 4SEASONS, and 16 pieces weighed about 4.2~g in total for the experiment on MERLIN. The single crystals on the plates were mounted in the ($H$,\,0,\,$L$) plane, with the [010] direction aligned in the vertical direction. The assembly was loaded into a closed-cycle refrigerator. Both 4SEASONS and MERLIN were operated in a multiple-$E_i$ (incident energy) mode. For the experiment on 4SEASONS, we used a set of $E_i=100.1$ (primary), 47.8, and 18.3~meV and a chopper frequency of 300~Hz. We set the angle of the neutron beam direction parallel to the $c$ axis to be zero. Data were collected at 6~K by rotating the sample about the [010] direction with a range of 60$^{\circ}$ in a 0.3$^{\circ}$ step. On MERLIN, data were collected in the range of 90$^{\circ}$ in a 1$^{\circ}$ step at 6~K, with a set of $E_i=90.2$ (primary), 31.8, and 16.0~meV and a chopper frequency of 300~Hz. We also conducted neutron scattering measurements at elevated temperatures on both spectrometers to investigate the temperature evolution of magnetic Bragg peaks. The wave vector ${\bm Q}$ was described by ($HKL$) reciprocal lattice unit (rlu) of $(a^{*},\,b^{*},\,c^{*})=(2\pi/a,\,2\pi/b,\,2\pi/c)$. Although there is a small orthorhombicity suggested in Ref.~\onlinecite{nm16_905}, we used a tetragonal notation with $a\approx b=4.4342$~\AA, and $c=23.1359$~\AA. Note that we used the longest axis as $c$, which is the same as that in Refs~\onlinecite{PhysRevB.95.134405,PhysRevB.100.144431} but different from that in Refs~\onlinecite{nm16_905,PhysRevB.100.205105}. In this case, our (001) corresponds to their (100).

\section{Results}

\subsection{Sample Characterizations}

We checked the orientation of the \sms single crystals with single-crystal and backscattering Laue X-ray diffractometers, and the results are shown in Fig.~\ref{fig1}(c) and (d), respectively. As shown in Fig.~\ref{fig1}(a), the crystal has a layered structure with Sb and Mn atoms forming square nets\cite{nm16_905,doi:10.1146/annurev-matsci-070218-010114}. The crystal naturally cleaves along the $c$ axis, so the cleavage plane can be nicely indexed with (0,\,0,\,$L$) as shown in Fig.~\ref{fig1}(c). The sharp (0,\,0,\,$L$) peaks in Fig.~\ref{fig1}(c) and clear Laue pattern in Fig.~\ref{fig1}(d) demonstrate the excellent crystallinity of the single crystals. EDS measurements on the single crystals show that there are approximately 1$\%$ Sr and 7$\%$ Mn deficiencies ($x<0.1$, $y<0.1$), which is commonly seen in this material\cite{nm16_905,PhysRevB.100.205105}. In fact, the magnetic properties of \sms are influenced by the amount of deficiency. Samples with larger Sr and smaller Mn deficiencies display stronger ferromagnetic behavior, while those with larger Mn deficiency show weaker ferromagnetic behavior\cite{nm16_905}. According to the magnitude of the ferromagnetic saturated moment $M_s$, Ref.~\onlinecite{nm16_905} categorized the samples into three types; $M_s\sim0.1-0.6~\mu_B/$Mn for type A, $\sim0.04-0.06~\mu_B/$Mn for type B, and $\sim0.004-0.006~\mu_B/$Mn for type C. Our samples with larger Mn deficiency and weaker ferromagnetic behavior fall into the category of type C. The temperature dependence of the magnetization is shown in Fig.~\ref{fig1}(e). At the $T_{\rm N}$ of $\sim$304~K, the magnetization starts to show a suppression due to the development of the antiferromagnetic order. This value is consistent with previous reports for this material\cite{nm16_905,PhysRevB.100.205105}. As shown in Fig.~\ref{fig1}(a), the antiferromagnetic order has a canted ferromagnetic moment along the $a$ axis\cite{nm16_905}. At low temperatures, the ferromagnetic moment develops and gives rise to the magnetization upturn shown in Fig.~\ref{fig1}(e). Such an upturn has also been observed in Ref.~\onlinecite{nm16_905}. The magnitude of the upturn corresponds to the strength of the ferromagnetic moment and depends on the stoichiometry, especially the amount of the Mn deficiency of the sample\cite{nm16_905}. Therefore, we suspect that the kink around 40~K in Fig.~\ref{fig1}(e) is due to the presence of a small amount of second phase with a slightly different stoichiometry. In Fig.~\ref{fig1}(f), we plot the temperature dependence of the integrated intensities of the magnetic Bragg peak (1,\,0,\,0), from which we observe that the peak intensity has an onset at the $T_{\rm N}$ of 304~K upon cooling. These results are consistent with previous literatures on this material\cite{nm16_905,PhysRevB.100.205105}.

\subsection{Topological Electronic Properties}

\begin{figure}[ht]
\centering
\includegraphics[width=0.95\linewidth]{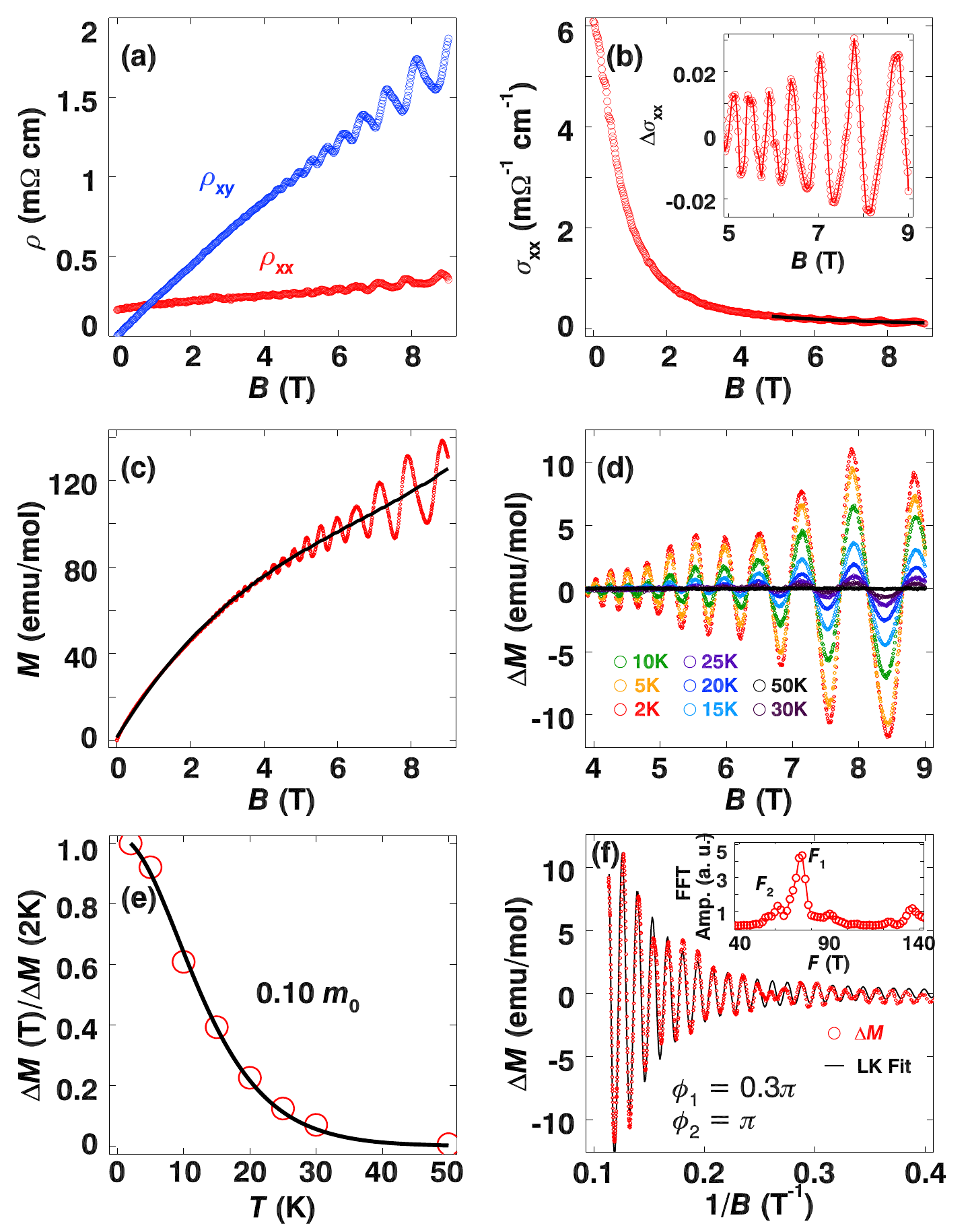}
\caption{(a) The in-plane longitudinal ($\rho_{xx}$) and transverse (Hall) ($\rho_{xy}$) resistivity as a function of magnetic field applied along the  $c$ axis measured at 2~K. (b) Magnetic-field dependence of the in-plane conductivity ($\sigma_{xx}$). The solid line indicates the smooth background. The inset is the oscillatory component ($\Delta\sigma_{xx}$) by subtracting the smooth background. (c) Magnetic-field dependence of the magnetization measured at 2~K, with the solid line being the fit of the smooth background. (d) Oscillatory component of the magnetization ($\Delta M$) by subtracting the background at various temperatures. (e) Temperature dependence of the oscillatory amplitude, normalized by $\Delta M$ at 2~K. The solid line through data is the fit as described in the main text, from which we obtain an effective mass of 0.1$m_0$. (f) $\Delta M$ at 2~K plotted as a function of $1/B$. The inset shows the fast Fourier transform spectra of $\Delta M$ which show two oscillating frequencies. The solid line is the fit to $\Delta M$ with a two-frequency Lifshitz-Kosevich formula. From the fit, we obtain the Berry phase of $0.3\pi$ and $\pi$ corresponding to the frequency of 73.2 and 57.2~T, respectively.
\label{fig2}}
\end{figure}

We first probed the electronic properties of \sms by measuring the resistivity and magnetization under external magnetic field. In Fig.~\ref{fig2}(a), we plot the in-plane longitudinal ($\rho_{xx}$) and transverse (Hall) ($\rho_{xy}$) resistivity measured at 2~K with field applied along the $c$ axis. In Fig.~\ref{fig2}(b), we plot the in-plane conductivity $\sigma_{xx}$, which is converted from the $\rho_{xx}$ and $\rho_{xy}$, using $\sigma_{xx}=\rho_{xx}/(\rho_{xx}^2+\rho_{xy}^2)$. By subtracting the background, we obtain the oscillatory component $\Delta\sigma_{xx}$, as shown in the inset of Fig.~\ref{fig2}(b). These strong SdH oscillations at moderate magnetic field suggest the high mobility and light effective mass of the carriers\cite{RevModPhys.81.109,Liang2015,Shekhar2015}. Furthermore, the large SdH oscillation amplitude is also an indication of the quasi-two-dimensional electronic structure of the material\cite{2004cr0306891}. In this study, we focus on the type-C samples. Compared to type-A and type-B samples which have larger saturated ferromagnetic moments\cite{nm16_905}, type-C samples exhibit relatively weak quantum oscillations due to the large $\rho_{xy}$. Moreover, SdH oscillations originate from the oscillating scattering rate and the details depend on the actual scattering mechanisms. This sometimes makes the Lifshitz-Kosevich (LK) theory not accurate in describing experimental observations\cite{2004cr0306891}, as is the case for \sms~(Ref.~\onlinecite{nm16_905}). On the other hand, the dHvA effect is caused directly by the oscillations of electrons' free energy and can be well described by the LK model for both three- and two-dimensional systems\cite{2004cr0306891}. Among the three types of samples, type C exhibits the strongest dHvA oscillations due to the weak ferromagnetic background\cite{nm16_905}. As shown in Fig.~\ref{fig2}(c), the magnetization shows very clear oscillations at moderate fields. Therefore, we use dHvA oscillations for detailed analyses of the electronic properties for our type-C samples below.

By subtracting the background of the magnetization, we obtain the oscillatory component of the magnetization $\Delta M$ as a function of magnetic field $B$~[Fig.~\ref{fig2}(d)]. The oscillatory component can be described by the LK formula\cite{2004cr0306891,PhysRevLett.82.2147}, as $\Delta M\propto~-B^{-\lambda}R_TR_DR_S\sin[2\pi(F/B-\gamma+\delta)]$, where the exponent $\lambda$ is 1/2 and 0 for the three- and two-dimensional cases, respectively;  Onsager phase factor $\gamma=1/2-\phi/2\pi$, and $\delta$ is the phase shift, taking the value of $\pm1/8$ and 0 for for three- and two-dimensional cases, respectively; $F$ and $\phi$ are the oscillation frequency and Berry phase, respectively; $R_{\rm T}=(2\pi^{2}k_{\rm B}T/\hbar\omega_{c})/\sinh(2\pi^{2}k_{\rm B}T/\hbar\omega_{c})$, and $R_{\rm D}=\exp(-2\pi^{2}k_{\rm B}T_{\rm D}/\hbar\omega_{c})$, with $\omega_{c}$ and $T_{\rm D}$ being the cyclotron frequency and Dingle temperature, respectively. $R_{\rm S}=\cos(\frac{1}{2}\pi gm^*/m_0)$ is the spin damping factor due to the Zeeman splitting, which can be neglected in the present case, as we do not observe any sign of Landau level splitting in the oscillations at $T=2$~K under $B=9$~T. Here, $m^*$ and $m_0$ are the effective mass and rest mass of the free electrons respectively.

Since the cyclotron frequency $\omega_{c}=eB/m^{*}$, the only temperature-dependent term in $\Delta M$ can be written as,
\begin{equation*}
\begin{split}
 R_T=(2\pi^2k_BT/\hbar\omega_c)/\sinh(2\pi^2k_BT/\hbar\omega_c)\\
 \approx(14.69m^*T/B)/\sinh(14.69m^*T/B).
\end{split}
\end{equation*}
We have fitted the temperature dependence of the amplitude of $\Delta M$ using this formula, and the results are shown in Fig.~\ref{fig2}(e). We obtain an effective mass of 0.10(3)$m_0$ for the sample by averaging the results obtained at several fields. This value is similar to those of NbP~(Ref.~\onlinecite{PhysRevB.93.121112}), ZrSiS~(\onlinecite{Singha2468}), and Cd$_3$As$_2$~(\onlinecite{PhysRevLett.114.117201,PhysRevLett.113.246402,PhysRevX.5.031037}). This effective mass is slightly different from that in Ref.~\onlinecite{nm16_905}, which may be caused by the composition difference and the limited magnetic field range in our study.

To obtain the Berry phase $\phi$ of \sms, we can use the Onsager quantization rule to construct the Landau-level fan diagram\cite{murakawa2013detection,doi:10.7566/JPSJ.82.102001}. However, as observed in Fig.~\ref{fig2}(d), there appear to be two oscillating frequencies. We perform fast Fourier transform analysis of the dHvA oscillation as shown in the inset of Fig.~\ref{fig2}(f). This clearly shows two frequencies of $F_1\approx73.2~$T and $F_2\approx57.2~$T, similar to the values for the type-C samples in Ref.~\onlinecite{nm16_905}. These values are in excellent agreement with the angle-resolved photoemission spectroscopy data and band structure calculations for this material\cite{SciPostPhys.4.2.010,Farhan_2014}. We further analyze the dHvA data with a two-frequency LK model, and the results are shown in Fig.~\ref{fig2}(f). From the LK fit, we obtain the Berry phase of 0.3$\pi$ and $\pi$ corresponding to the bands of 73.2 and 57.2~T, respectively. These strongly indicate that the band structure is topologically nontrivial. We have measured the resistivity and magnetizations for other type-C samples. Although the details depend on the stoichiometry of the samples, all of them exhibit nontrivial topological properties with nonzero Berry phase and light effective mass. These results are consistent with the suggestion that type-A, -B, and -C samples are all topologically nontrivial despite of the quantitative difference among different sample types\cite{nm16_905}.

\subsection{INS Spectra}

\begin{figure*}[ht]
\centering
\includegraphics[width=0.95\linewidth]{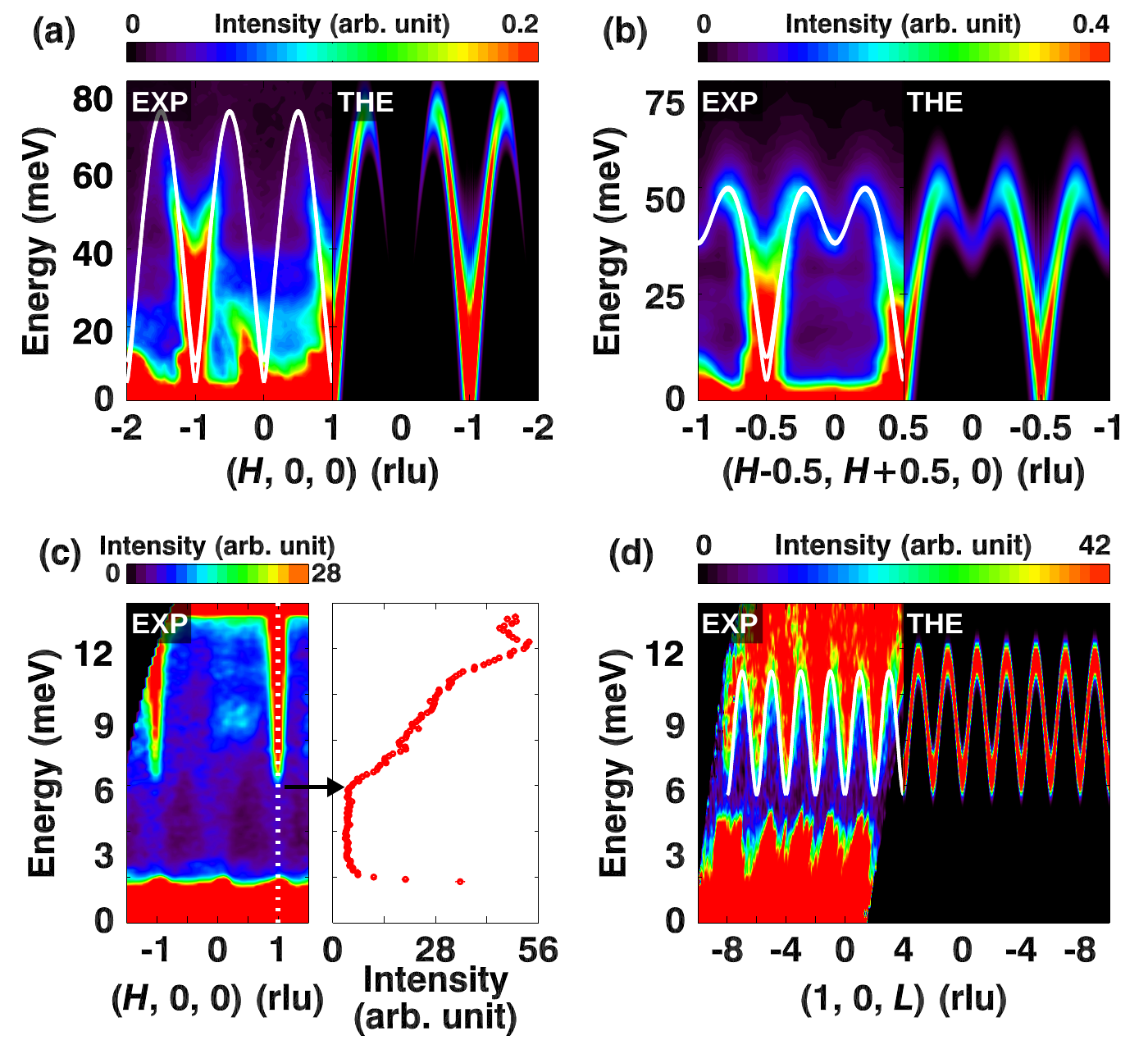}
\caption{Magnetic dispersions of \sms measured at 6~K together with the theoretical calculation results. (a) and (b) Dispersions along the [100] and [110] directions, respectively, obtained on 4SEASONS with $E_i=100.1$~meV. (c) and (d) Dispersions along the [100] and [001] directions obtained on MERLIN with $E_i=16$~meV. Panel (c) focuses on the low-energy part of (a) to show the spin gap. In panels (a), (b), and (d), solid lines are the calculated dispersions. The calculated dispersions with intensities are shown on the right hand side of each panel to be compared with the experimental data on the left hand side. Data in (a) and (c) were integrated with $K=[-0.2,0.2]$~rlu, and all available $L$s; data in (b) with a thickness of 0.1~rlu about each momentum along the [-110] direction; data in (d) with $H=[0.8,1.2]$~rlu, and $K=[-0.2,0.2]$~rlu. The dashed line in (c) illustrates an energy scan at (100). The scan profile is plotted on the right hand side with the energy and intensity being the $y$ and $x$ axis, respectively. The arrow indicates the gap value. In (d), the low-energy part below about 4.5~meV is due to the Bragg peaks and some spurious signals, and the broad experimental data can be reproduced by spin-wave calculations convoluting the instrumental resolution\cite{PhysRevB.95.134405}.
\label{fig3}}
\end{figure*}

\begin{figure*}[ht]
\centering
\includegraphics[width=0.95\linewidth]{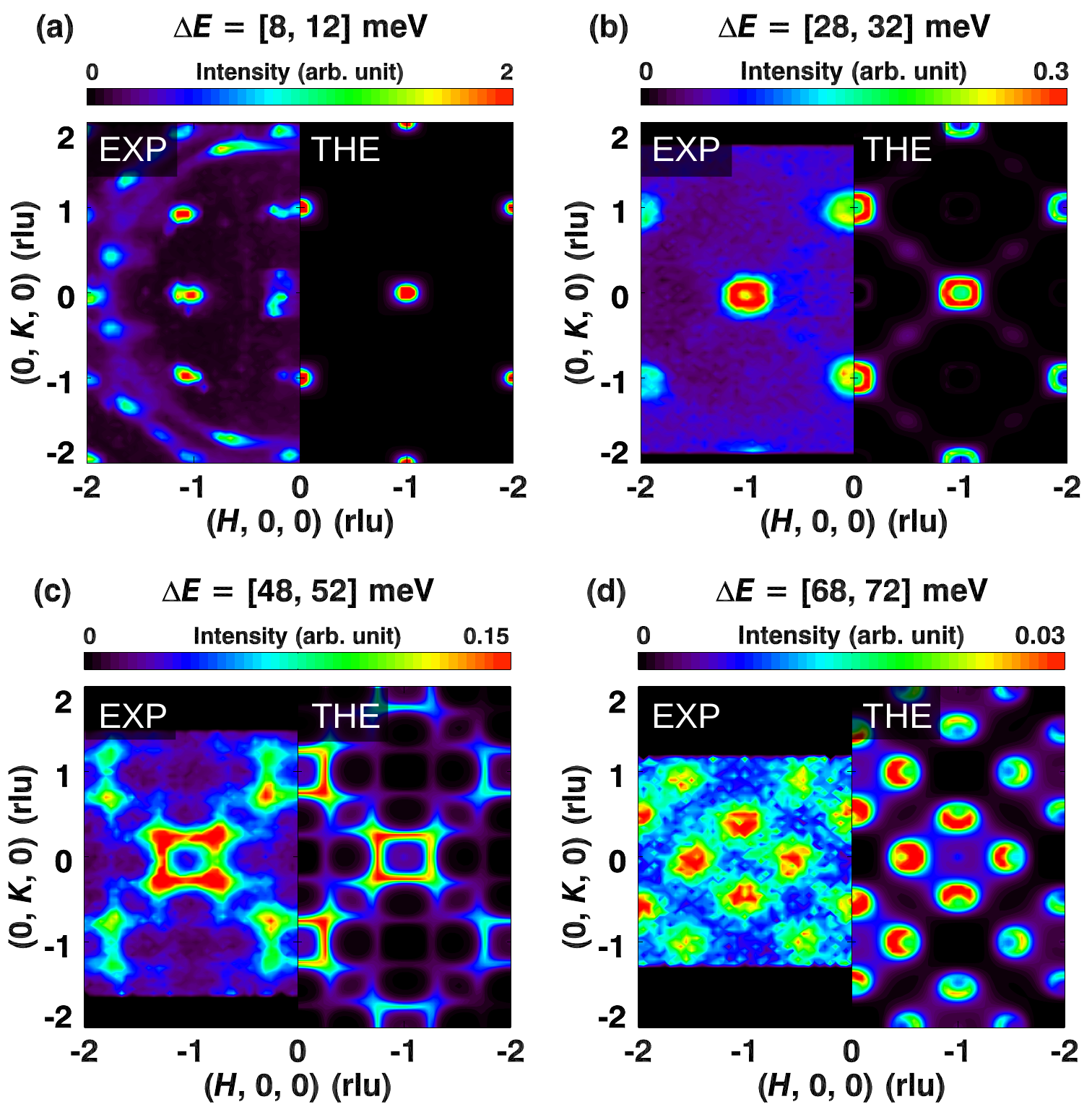}
\caption{Contour maps of the constant-energy cuts measured with $E_i=100.1$~meV at 6~K. The experimental data are shown on the left hand side of each panel, while the theoretical results are shown on the right hand side. The energy range of the integration is shown on the top of each panel. All the data are from 4SEASONS.
\label{fig4}}
\end{figure*}

Having studied the topological electronic properties of the material, we now move to investigate the spin dynamics by INS measurements to explore whether and how topological fermions affect magnons. In Fig.~\ref{fig3}(a) and (b), we present magnetic dispersions along two high-symmetry directions of [100] and [110], respectively, measured with an incident energy $E_i$ of 100.1~meV at 6~K. Along the [100] direction, the excitation spectra disperse from the antiferromagnetic ordering wavevector (1,\,0,\,0) and reach the band top at about 76~meV at (0.5,\,0,\,0) and (-1.5,\,0,\,0) [Fig.~\ref{fig3}(a)]. Note that the (1,\,0,\,0) peak is structurally forbidden\cite{nm16_905}, and only appears below the antiferromagnetic transition temperature $T_{\rm N}$ of 304~K as shown in Fig.~\ref{fig1}(f), so these excitations are not contributed by phonons but purely magnetic. The weak spectral weight for wave vectors $\bm{Q}$s ranging from (-0.5,\,0,\,0) to (0.5,\,0,\,0) is due to the small magnetic structure factor. The signal at the $\Gamma$ point with $H=0$ is expected to be weak due to the antiferromagnetic structure\cite{nm16_905}. The spin waves remain well defined up to about 70~meV. Along the [110] direction, the band top is reduced to about 50~meV, as shown in Fig.~\ref{fig3}(b), indicating the strong anisotropy within the $a$-$b$ plane. The dispersion shows a local minimum at (-0.5,\,0.5,\,0). To probe the low-energy excitations, we used an $E_i$ of 16~meV, and the obtained dispersion along the [100] direction is shown in Fig.~\ref{fig3}(c). The dispersion in the left panel of Fig.~\ref{fig3}(c) clearly shows a spin gap of 6~meV. The gap is also evident from the cut at (100) along the energy axis [right panel of Fig.~\ref{fig3}(c)]. Although the spin-gap value is similar to that obtained for the type-A sample with strong ferromagnetic moment in Ref.~\onlinecite{PhysRevB.100.205105}, the high-energy behavior in Fig.~\ref{fig3}(c) is different from that in Ref.~\onlinecite{PhysRevB.100.205105}, possibly because we use type-C samples, different from Ref.~\onlinecite{PhysRevB.100.205105}. In Fig.~\ref{fig3}(d), we plot the dispersion along the [001] direction. The out-of-plane dispersion also exhibits a clear spin gap at 6~meV, and the total band width is about 5~meV, an order of magnitude smaller than those in the plane shown in Fig.~\ref{fig3}(a) and (b), consistent with the quasi-two-dimensional crystal structure of this material. The magnetic excitations with a spin gap of 6~meV, highly dispersive in the plane, but weakly dispersive out of the plane, are qualitatively similar to those in related magnetic Dirac materials such as CaMnBi$_2$\cite{PhysRevB.95.134405}, SrMnBi$_2$\cite{PhysRevB.95.134405}, and YbMnBi$_2$\cite{PhysRevB.100.144431,arXiv:1908.08114}. These spin-wave excitations remain well defined at high energies, distinct from some iron-based superconductors where the high-energy excitations are strongly damped and the spin waves decay into a Stoner continuum\cite{PhysRevLett.102.187206,PhysRevB.83.214519}.

\begin{table*}[htb]
  \begin{threeparttable}
\caption{Exchange parameters, single-ion anisotropy constants, spin gaps, ordering temperature $T_N$, and related references. Values in the brackets denote the errorss.}
\label{tab:para}
\begin{tabular*}{\textwidth}{@{\extracolsep{\fill}}cccccccccccccc}
\hline \hline
\begin{minipage}{2cm}\vspace{1mm}   \vspace{1mm} \end{minipage} & $SJ_1$ (meV) & $SJ_2$(meV) & $SJ_C$(meV) & $SD$(meV) & $\Delta$(meV) & $T_N$ (K) & References\\
\hline
\begin{minipage}{2.5cm}\vspace{1mm}~\sms\vspace{1mm} \end{minipage} & 28.0(2) & 9.27(36) & -0.10(1) & -0.07(8) & $\sim$6.0 & $\sim$304 & our work\\ \sms
   & 23.25(75) & negligible & -0.65(20) & diag(0.30(13) 0.8(3) 0)* & $\sim$8.5 & $\sim$304 & \onlinecite{PhysRevB.100.205105}\\SrMnBi$_2$ & 21.3(2) & 6.39(15) & 0.11(2) & 0.31(2) & 10.2(2) & 287(5) & \onlinecite{PhysRevB.95.134405}\\CaMnBi$_2$ & 23.4(2) & 7.9(5) & -0.10(5) & 0.18(3)& 8.3(8) & 264(2) & \onlinecite{PhysRevB.95.134405}\\YbMnBi$_2$ & 25.2(2) & 10.1(3) & 0.131(4) & -0.20(1) & 9.0(2) & $\sim$290 & \onlinecite{PhysRevB.100.144431,arXiv:1908.08114}\\
\hline \hline
\end{tabular*}
\begin{tablenotes}
      \small
      \item \quad *In Ref.~\onlinecite{PhysRevB.100.205105}, $S^{\alpha}A_{i}^{\alpha\beta}S^{\beta}$ was used to replace the $SD$ in the Hamiltonian.\\

\end{tablenotes}
\end{threeparttable}
\end{table*}

To visualize the magnetic excitations in the momentum space, we plot a series of constant-energy contours in the ($HK$0) plane with energies up to 70~meV in Fig.~\ref{fig4}. At an energy transfer of $\Delta E=10\pm2$~meV, slightly above the spin gap of 6~meV, there are both magnetic and lattice excitations. For instance, in Fig.~\ref{fig4}(a), there is a strong peak at (-1,\,0,\,0), which corresponds to a magnetic Bragg peak at the elastic position. On the other hand, the peaks at (-1,\,$\pm1$,\,0) are phonons, which disappear at higher energies, as the phonon intensities decrease rapidly as the energy increases. The spin waves disperse outward as the energy increases and form a ring centering (-1,\,0,\,0) as shown in Fig.~\ref{fig4}(b); upon further increasing the energy to 50~meV, the ring evolves into a square. At 70~meV, slightly below the band top, there are still well-resolved magnetic excitations. The clear patterns of the spin waves allow us to obtain an accurate parameterization of the spectra, as we discuss below.

To analyze the data, we use the following Hamiltonian\cite{PhysRevB.95.134405,PhysRevB.100.144431,arXiv:1908.08114}
\begin{equation*}
\hat{H}=\sum_{<\emph{i},\emph{j}>}\emph{J}_\emph{ij}\hat{S_i}\cdot\hat{S_j}-\sum_{\emph{i}}\emph{D}(\hat{S_i^Z})^2,
\end{equation*}
where we include a nearest-neighbor ($J_1$) and next-nearest-neighbor ($J_2$), and interlayer exchange coupling constant ($J_c$). The exchange paths are illustrated in Fig.~\ref{fig1}(a). To account for the spin gap, we include a single-ion anisotropy term $D$. Based on the linear spin-wave theory and SPINW package\cite{0953-8984-27-16-166002}, we calculated the spin-spin correlation function\cite{neutron2,neutron1},
\begin{equation*}\label{SS}
   S^{\alpha\beta}(\bm{Q},\omega)=\frac{1}{N}\sum_{ij}{\rm e}^{\mathrm{i}\bm{Q}(\bm{r_i}-\bm{r_j})}\int_{-\infty}^{\infty}\left\langle S^\alpha_{i} S^\beta_{j}(t)\right\rangle {\rm e}^{-\mathrm{i}\omega t}dt,
\end{equation*}
which is related to scattering cross section
\begin{equation*}\label{SS1}
   \frac{d^2\sigma}{d\Omega d\omega}\propto|F(\bm{Q})|^2\sum_{\alpha\beta}(\delta_{\alpha\beta}-\frac{Q_\alpha Q_\beta}{Q^2})S^{\alpha\beta}(\bm{Q},\omega),
\end{equation*}
where $\alpha$ and $\beta$ are direction indices ($xyz$), $F(\bm{Q})$ is the magnetic structure factor, $\bm{S}_{i}$ is the spin at site $i$ with the coordinate $\bm{r}_i$, and $N$ is the total number of sites. In Figs~\ref{fig3} and \ref{fig4}, the calculated spectra using $SJ_1\sim28.0$, $SJ_2\sim9.3$, $SJ_c\sim-0.1$, and $SD\sim-0.07$~meV are plotted on the right hand side of the experimental results in Fig.~\ref{fig3}(a), (b), and (d). Furthermore, the calculated dispersions are plotted as solid lines on top of the experimental dispersions. It is clear that the calculations are in excellent agreement with the experimental data, including the dispersions in and out of plane, the spin anisotropy gap, and the spectral weight distribution in the momentum space. The parameters we used in these calculations, and the gap as well as the $T_{\rm N}$ are tabulated in Table~\ref{tab:para} together with data for other related materials.

There are a few remarks we want to make about the calculations and fittings. First, since there are four Mn atoms in each unit cell\cite{nm16_905}, there are four spin-wave branches in the calculations as shown in Fig.~\ref{fig3}(a) and (b). However, due to the resolution limit, we are not able to resolve these four branches experimentally. Second, the structure factor is small for $\bm{Q}$ ranging from (-0.5,\,0,\,0) to (0.5,\,0,\,0), consistent with the experimental observations of vanishing scattering intensities in this range. Third, a positive $J_2$
is needed to account for the local minimum at (-0.5,\,0.5,\,0), as shown in Fig.~\ref{fig3}(b). The perfect match of the calculated dispersion and the experimental spectra further demonstrates the quality of the fits. Without such a $J_2$, the local minimum is absent, as shown in Ref.~\onlinecite{PhysRevB.100.205105}. The same sign of $J_1$ and $J_2$ introduces frustration into the system, as is also the case for CaMnBi$_2$\cite{PhysRevB.95.134405}, SrMnBi$_2$\cite{PhysRevB.95.134405}, and YbMnBi$_2$\cite{PhysRevB.100.144431,arXiv:1908.08114}, as tabulated in Table~\ref{tab:para}. Fourth, the spin gap can be well reproduced by including the single-ion anisotropy term. In Ref.~\onlinecite{nc7_13833}, by fitting the two-magnon Raman spectra in CaMnBi$_2$ and SrMnBi$_2$, an anomalously large interlayer coupling $J_c$ was obtained for both compounds and interpreted as due to the Bi Dirac band structure. In that work, the single-ion anisotropy term was neglected. In our study, we find that the $J_c$ is fairly small ($SJ_c\sim-0.10$~meV), similar to other INS work on related materials\cite{PhysRevB.95.134405,PhysRevB.100.144431,arXiv:1908.08114}. Such a small $J_c$ is consistent with the quasi-two-dimensional structure of these materials. Fifth, by comparing the experimental and theoretical results in Fig.~\ref{fig4}(a), it is clear that the spin waves are well separated from the phonons as they emerge from different Bragg peaks. This makes analyzing the spin waves more easily. Sixth, the calculations of the dispersion along the [001] direction are much sharper than the experimental data, as shown in Fig.~\ref{fig3}(d). Similar behavior was also observed in Ref.~\onlinecite{PhysRevB.95.134405}, and it was shown that by convoluting the instrumental resolution in the calculations, the broad experimental spectra could be reproduced.

\section{Discussions and Conclusions}

From the electrical transport and magnetization results, we find that \sms hosts
topological fermions, consistent with the proposal of magnetic Weyl semimetal state in this system\cite{nm16_905}. On the other hand, we observe well-defined spin waves resulting from the Mn atoms, with no discernible damping due to the itinerant electrons. The spectra can be well fitted by a local-moment Heisenberg model. From these data, we do not detect any effect of the itinerant electrons on the spin-wave excitations resulting from the Mn local moments. Similarly, in CaMnBi$_2$\cite{PhysRevB.95.134405}, SrMnBi$_2$\cite{PhysRevB.95.134405}, and YbMnBi$_2$\cite{PhysRevB.100.144431}, no influence of the Dirac fermions on the spin dynamics has been observed either. These behaviors are different from those in iron-based superconductors where itinerant electrons are believed to play an important role in damping the high-energy magnetic excitations and making the spin waves decay into a Stoner continuum\cite{PhysRevLett.102.187206,PhysRevB.83.214519}.

In \sms, the magnetic excitations are resulting from the local moments in the Mn layers, whereas the itinerant electrons are contributed by the Sb layers which are separated from the former\cite{PhysRevLett.107.126402,SciPostPhys.4.2.010,Farhan_2014,PhysRevB.95.134405,PhysRevB.100.144431}. The material has a quasi-two-dimensional structure with a large $c$-axis constant of 23.1359~\AA. The large SdH oscillation amplitude also indicates that the electronic structure for \sms is two-dimensional like.
Furthermore, the spin-wave spectra indicate that the magnons are quasi-two-dimensional as well. Our fitting indicates that $SJ_1$ and $SJ_2$ are 28.0 and 9.27~meV, respectively, while $J_c$ is only -0.10~meV, two orders of magnitude smaller. Taking these into account, it is not surprising that we do not observe the effect of the itinerant electrons on the spin dynamics. We believe this also explains the similar behavior in other magnetic Dirac semimetal systems\cite{PhysRevB.95.134405,PhysRevB.100.144431}. Nevertheless, our results do not completely rule out the possible subtle influence of the itinerant electrons on the magnons.

Baring these in mind, in order to observe strong spin-fermion coupling, it may be helpful to look for quasi-two-dimensional magnetic Dirac or Weyl semimetals with stronger interlayer couplings, or even three-dimensional materials. A more ideal case will be that the magnons and fermions are in the same layer or arising from the same atoms like in iron-based superconductors\cite{PhysRevLett.102.187206,PhysRevB.83.214519}. The recently discovered magnetic Weyl semimetal Co$_3$Sn$_2$S$_2$ may be such an example\cite{Morali1286,Liu1282}. In this material, it was shown that the Weyl nodes were close to the Fermi level\cite{PhysRevB.97.235416,Liu1282}. Furthermore, it exhibited both high anomalous Hall effect and large anomalous Hall angle\cite{np14_1125,nc9_3681}, indicating strong coupling between the magnetism and electronic transport. It would be of great interest to further explore the possible spin-fermion coupling in this and other related materials\cite{Belopolski1278}.

Likewise, to investigate the spin-fermion interactions, one can also make discussions in the context of how the electronic structures are affected by the magnetism. For example, in a magnetic Weyl semimetal Mn$_3$Sn, it was suggested that the helical magnetic ordering observed in the INS measurements could remove the Weyl nodes in the helical phase\cite{npj3_63}. Similar practice had also been taken in \sms, where the authors extracted the exchange parameters from the spin excitation spectra obtained from INS measurements, and then with the aid of density functional theory, they suggested that the magnetic order could tune the topological band structure\cite{PhysRevB.100.205105}. However, it is not clear how the inclusion of a substantial $J_2$, as shown in our current work and typically present in these materials (Table~\ref{tab:para}), will affect the results. It will be beneficial to carry out angle-resolved photoemission spectroscopy measurements to test these proposals.

To conclude, while our electrical transport and magnetization measurements indicate that \sms is a topological semimetal, our INS data reveal well-defined spin waves resulting from the Mn local moments, which can be nicely described using a Hamiltonian including $SJ_1\sim28.0$~meV, $SJ_2\sim9.3$~meV, $SJ_c\sim-0.1$~meV, and $SD\sim-0.07$~meV. Our results suggest that the topological fermions do not impact the magnons. We propose this to be due to the separated layers of the localized spins and itinerant carriers and the weak interlayer coupling.

\section{Acknowledgements}

We would like to thank Yuefeng Nie for providing access to the single-crystal X-ray diffractometer. We are grateful for the stimulating discussions with Fucong Fei. The work was supported by the National Natural Science Foundation of China with Grant Nos~11822405, 11674157, 11674158, 11774152, and 11904170, National Key
Projects for Research and Development of China with Grant No.~2016YFA0300401, Natural Science Foundation of Jiangsu Province with Grant Nos~BK20180006 and BK20190436, Fundamental Research Funds for the Central Universities with Grant No.~020414380117, and the Office of International Cooperation and Exchanges of Nanjing University. The experiment at the Materials and Life Science Experimental Facility of the J-PARC was performed under a user program (Proposal No.~2018A0018). We also acknowledge the financial support by the Newton Fund for China from ISIS Facility of Rutherford Appleton Laboratory.



%

\end{document}